\documentclass[a4paper,10pt]{article} 

\usepackage[utf8]{inputenc}
\usepackage[T1]{fontenc}
\usepackage[frenchb,english]{babel}
\usepackage{textcomp}
\usepackage{amsmath}
\usepackage{amsfonts}
\usepackage{amssymb}
\usepackage{graphicx}
\usepackage{vmargin}
\usepackage{url}
\usepackage[bookmarks=true,bookmarksnumbered=true,colorlinks=false,pdfborder={0 0 0}]{hyperref}
\usepackage{fancyhdr}
\usepackage{lastpage}
\usepackage{natbib}

\setmarginsrb{25mm}{20mm}{25mm}{15mm}{12pt}{5mm}{0pt}{11mm}
\newcommand{\ve}{\mathbf}

\begin{document}


\pagestyle{fancy}
\chead{}
\lhead{\textit{B. Lenoir et al., 2011}}
\rhead{\textit{\thepage/\pageref{LastPage}}}
\cfoot{}
\renewcommand{\headrulewidth}{0pt}
\renewcommand{\footrulewidth}{0pt}



\title{Electrostatic accelerometer with bias rejection for gravitation \\ and Solar System physics}
\author{Benjamin Lenoir\textsuperscript{a}, Agnès Lévy\textsuperscript{a}, Bernard Foulon\textsuperscript{a}, Brahim Lamine\textsuperscript{b}, Bruno Christophe\textsuperscript{a}, \\ Serge Reynaud\textsuperscript{b} \\ \small \textsuperscript{a} \textit{Onera -- The French Aerospace Lab, 29 avenue de la Division Leclerc, F-92322 Ch\^atillon, France} \\ \small \textsuperscript{b} \textit{Laboratoire Kastler Brossel (LKB), ENS, UPMC, CNRS, Campus Jussieu, F-75252 Paris Cedex 05, France} \\ \\ \small Published in \textit{Advances in Space Research} 48 (2011) 1248-1257}
\date{23 August 2011}
\maketitle



\begin{abstract}
Radio tracking of interplanetary probes is an important tool for navigation purposes as well as for testing the laws of physics or exploring planetary environments. The addition of an accelerometer on board a spacecraft provides orbit determination specialists and physicists with an additional observable of great interest: it measures the value of the non-gravitational acceleration acting on the spacecraft, i.e. the departure of the probe from geodesic motion.

This technology is now routinely used for geodesy missions in Earth orbits with electrostatic accelerometers. The present article proposes a technological evolution which consists in adding a subsystem to remove the measurement bias of an electrostatic accelerometer. It aims at enhancing the scientific return of interplanetary missions in the Solar System, from the point of view of fundamental physics as well as Solar System physics. The main part of the instrument is an electrostatic accelerometer called MicroSTAR, which inherits mature technologies based on Onera's experience in the field of accelerometry. This accelerometer is mounted on a rotating stage, called Bias Rejection System, which modulates the non-gravitational acceleration and thus permits to remove the measurement bias of the instrument from the signal of interest.

The article presents the motivations of this study, describes the instrument, called GAP, and the measurement principle, and discusses the performance of the instrument as well as integration constraints. Within a total mass of 3.1 kg and an average consumption of 3 W, it is possible to reach an accuracy of 1 pm s$^{-2}$ for the acceleration measured with an integration time of five hours. Combining this observable with the radio tracking data, it is therefore possible to compare the motion of the spacecraft to theoretical predictions with an accuracy improved by at least three orders of magnitude with respect to existing techniques.


\end{abstract}



\section{Introduction}\label{section:introduction}

Gravitation is one of the four interactions in the standard model of fundamental physics. General Relativity, its current theoretical formulation, is in good agreement with most experimental tests of gravitation \citep{will2006confrontation}. But General Relativity is a classical theory and all attempts to merge it with the quantum description of the other fundamental interactions suggest that it is not the final theory for gravitation.

General Relativity is also challenged by observations at galactic and cosmic scales. The rotation curves of galaxies and the relation between redshifts and luminosities of supernovae deviate from the predictions of General Relativity. These anomalies are interpreted as revealing the presence of new components in the content of the Universe, the so-called ``dark matter'' and ``dark energy'' which are thought to constitute respectively 25\% and 70\% of the energy content of the Universe \citep{copeland2006dynamics,frieman2008dark}. Their nature remains unknown and, despite their prevalence, they have not yet been detected by any other means than gravitational measurements. Given the immense challenge posed by these large scale behaviors, it is necessary to explore every possible explanation including the hypothesis that General Relativity is not a correct description of gravitation at galactic and cosmic scales \citep{aguirre2001astrophysical,nojiri2007introduction}.

Testing gravitation at the largest scales reachable by man-made instruments is therefore essential to bridge the gap between experiments in the Solar System and astrophysical or cosmological observations. The most notable experiment was the test of the gravitation law performed by NASA during the extension of the Pioneer 10 \& 11 missions. This test resulted in what is now known as the ``Pioneer anomaly'' \citep{anderson1998indication,anderson2002study}, one of the few experimental signals deviating from the predictions of General Relativity \citep{lammerzahl2008physics,anderson2009astrometric}.

In a context dominated by the questions of dark matter and dark energy, the challenge raised by the anomalous Pioneer signals has to be faced. Efforts have been devoted to the reanalysis of Pioneer data \citep{markwardt2002independent,olsen2007constancy,levy2009pioneer,turyshev2009pioneer} with the aim of learning as much as possible on its possible origin, which can be an experimental artifact as well as a hint of considerable importance for fundamental physics \citep{reynaud2008tests,reynaud2009testing,turyshev2010pioneer}. In the meantime, theoretical studies have been devoted to determine whether or not the anomalous signal could reveal a scale-dependent modification of gravitation law while remaining compatible with other tests. Among the candidates are in particular metric extensions \citep{jaekel2005gravity,jaekel2005post,jaekel2006post,jaekel2006radar} as well as multifield extensions \citep{moffat2005gravitational,moffat2006scalar,bruneton2007field} of General Relativity.

Naturally, several mission concepts have been proposed \citep{anderson2002mission,dittus2005mission,johann2008exploring,bertolami2007mission,christophe2009odyssey,wolf2009quantum} to improve the experiment performed by Pioneer 10 \& 11 probes. A key idea in most of these proposals is to embark accelerometers so as to measure the non-gravitational forces acting on the spacecraft, whatever may be the cause of these forces. The addition of this observable not only improves the accuracy and quality of the navigation of the spacecraft but also allows understanding the origin of the Pioneer anomalous signals by measuring the non-gravitational acceleration of the spacecraft. Indeed, with such an instrument, no models are needed to estimate non-gravitational forces, which removes a source of uncertainty in the orbit determination process.

The main subsystem of the instrument is an electrostatic accelerometer called MicroSTAR, which inherits mature technologies. The main evolution of the instrument is to mount MicroSTAR on a rotating stage, called Bias Rejection System, which modulates the signal of interest, and thus makes it possible to remove the measurement bias of the electrostatic accelerometer. After a presentation of the motivations for this study, the accelerometer and the measurement principle will be described. The performance of the instrument will then be discussed. Finally, the integration constraints, which depend on the spacecraft design, will be briefly presented. A precision of 1 pm s$^{-2}$ for the acceleration measured with an integration time of five hours will be demonstrated. Using these measurements along with radio tracking data, the instrument enables to enhance by at least three orders of magnitude the accuracy of the comparison between the spacecraft gravitational acceleration and theoretical predictions.

The instrument studied in this article is an important technological upgrade for the future of fundamental physics in space as well as for the exploration of the outer Solar System. It may indeed be flown as a passenger on planetary missions heading toward outer planets. It will improve the science return with respect to objectives in fundamental physics -- deep space gravitation tests -- as well as in Solar System physics -- study of the gravity field and environment of the outer planets or their moons \citep{milani2001gravity}. This idea has been included in the Roadmap for Fundamental Physics in Space issued by ESA in 2010 \citep{esa2010roadmap}. Indeed, these two scientific goals are intimately connected since the gravitation force law is related to the ephemeris of planets, moons and minor objects \citep{fienga2010gravity} as well as to the origins of the Solar System \citep{blanc2005tracing}.

\section{Overview and motivations}\label{section:overview}

The trajectory of a spacecraft, which is tracked by the radio link \citep[e.g.][]{zarrouati1987trajectoires,moyer2000formulation}, is the result of gravitational and non-gravitational forces. The knowledge of the non-gravitational forces makes it possible to compute the geodesic motion of the spacecraft from the radio tracking data. This geodesic motion can then be compared to predictions of gravitation theories, allowing to test gravitation at the Solar System scale.

\subsection{Orbit reconstruction without accelerometric measurement}

The following analysis is made in a post-Newtonian framework which is sufficient for the purpose of this article. The tracking techniques give access to the Doppler acceleration of the spacecraft along the line of sight, $a_{Doppler}$. This Doppler acceleration includes the non-gravitational acceleration, $\ve{a_{NG}}$, and a term called $a_{other}$. This last term includes the acceleration of the spacecraft due to the curvature of space-time, the interaction of the radio waves with the gravitational field, the speed and acceleration of the Earth station tracking the spacecraft, and aberration effects. Calling $\ve{u}$ the unit vector directing the line of sight, this leads to the simplified formula:
\begin{equation}
  a_{Doppler} = \ve{a_{NG}}\cdot\ve{u} + a_{other}
\end{equation}
When $a_{Doppler}$ is the only observable, it is necessary to use models to correct for $\ve{a_{NG}}$ in order to access $a_{other}$, which can be compared to theoretical predictions. This approach has been implemented for all interplanetary spacecrafts' orbit determinations so far \citep{tortora2004precise,asmar2005spacecraft}. But the orbit determination is then subject to errors due to uncertainties or inaccuracies in the models. Even when the average non-gravitational acceleration is properly corrected, temporal fluctuations are poorly known.

\subsection{Orbit reconstruction with accelerometric measurement}

The electrostatic accelerometer with bias rejection GAP measures along two orthogonal axis (directed by $\ve{e_1}$ and $\ve{e_2}$) the absolute value of the non-gravitational acceleration acting on the spacecraft $\ve{a_{NG}}$. When $\ve{u}$ is in the plane generated by $\ve{e_1}$ and $\ve{e_2}$, one can retrieve the quantity $a_{other}$ by combining the Doppler and the accelerometer measurements.

This provides an additional observable which enhances the orbit reconstruction for several reasons. First, it removes parameters to be fitted in the process and consequently improves the quality of the orbit determination. Second, it measures the fluctuations of the non-gravitational acceleration, which cannot be taken into account by models. Third, it removes the correlations which appear in the orbit determination process between the non-gravitational acceleration and the gravitational acceleration, when the former are not measured. This approach has been used for CHAMP and GRACE geodesy missions \citep{touboul1999accelerometers}. In the frame of a test of the gravitation law in the Solar System, the presence of an accelerometer becomes essential in order to achieve the best possible accuracy on the orbit determination.

The instrument presented in this article has been designed for fundamental physics purposes, but it can also be used for orbit reconstruction around planets or during fly-bys \citep{christophe2008gravity}. In terms of planetary scientific objectives, the measurement of the non gravitational acceleration would improve the reconstruction of the gravity fields and therefore the knowledge of the body interior structure. It would also give information on the atmosphere with spatial and temporal resolutions outreaching by far those provided without accelerometric measurement.

\subsection{Sources of non-gravitational accelerations}\label{subsection:NGacc}

In this section, different sources of non-gravitational accelerations are discussed for an interplanetary spacecraft. Depending on the power requirement on-board and on the target distance from the Sun (among other considerations), the power generation can be made using solar panels or Radio-isotope Thermal Generator (RTG). In the first case, the predominant effect is the direct solar radiation pressure whereas, in the second one, it is the thermal radiation.

The direct solar radiation pressure corresponds to the action of solar photons on a unit surface. The power carried by solar photons by surface unit at one astronomical unit from the Sun, called $d_0$, is approximately equal to $P = 1.366 \times 10^3$ W m$^{-2}$ \citep{willson2003secular}. Considering a ballistic coefficient equals to $C_B = 0.1$ m$^2$ kg$^{-1}$, which is the order of magnitude for Laplace mission \citep{biesbroek2008laplace}, the acceleration due to radiation pressure is equal to $C_B P / c = 4.6 \times 10^{-7}$ m s$^{-2}$ at one astronomical unit, where $c$ is the speed of light. When the spacecraft is at a distance $d$ from the Sun, this value is multiplied by the $(d_0/d)^2$. The fluctuations of this acceleration have been characterized by \citet{frohlich2004solar} for frequencies between $10^{-8}$ Hz and $10^{-2}$ Hz.

The anisotropic thermal radiation of the spacecraft induces a force which is difficult to analyze and predict. One of the reasons is the time variation of the thermo-optics coefficients of the materials on the spacecraft and the difficulty to forecast them, as illustrated in the analysis by \citet{bertolami2008thermal}, \citet{toth2009thermal} or \citet{rievers2010modeling}. The use of an instrument like GAP removes any problem concerning the modeling of these forces since they are directly measured.

When a spacecraft orbits or fly-bys a planet, the atmosphere induces a force on the spacecraft which depends on the relative speed, the density and the ballistic coefficient. Models of atmospheric density exist but their accuracy has not been proved and they do not take into account any time variability. Therefore, computation of drag forces is plagued by large uncertainties.

The interplanetary medium also induces a drag force on the spacecraft during its cruise. This effect has been evaluated for the Pioneer probes and found to be small \citep{nieto2005directly}.

Finally the magnetic field acts on the spacecraft because of its charge via the Lorentz force. When close to a planet, the spacecraft will also experience the radiation pressure due to the planet's albedo. Other forces which are not listed here may also play a role (e.g. solar wind, maneuver, gas leakage).

\subsection{Electrostatic measurement}

The electrostatic accelerometer, MicroSTAR, measures a combination of the non-gravitational accelerations presented above and other terms described later in this article. Along one axis, this external signal, which is to be measured, is called $s$ and the actual measurement made by MicroSTAR is called $m$. The measurement chain induces a noise and a bias, respectively called $n$ and $b$, as well as linear and quadratic factors, respectively called $\delta k_1$ and $k_2$. The bias corresponds to the low-frequency fluctuations and drifts with respect to the duration of the measurement. The relation between these quantities is given by:
\begin{equation}
  m = (1+\delta k_1) s + k_2 s^2 + b + n
  \label{eq:bias_noise}
\end{equation}
In the following, $\delta k_1$ and $k_2$ will be assumed to be equal to zero without jeopardizing the discussion. The purpose of the instrument presented in this article is to remove the bias via a modulation of the external signal using a rotation of MicroSTAR with a rotating stage called Bias Rejection System. First, the electrostatic accelerometer MicroSTAR is presented, as well as the spectral characterization of its noise. Then, the measurement principle and the bias rejection scheme are described. This gives the opportunity to give some requirements on the Bias Rejection System needed to achieve the target accuracy.

\section{MicroSTAR: a miniaturized electrostatic accelerometer}\label{subsection:MicroSTAR}

MicroSTAR is an electrostatic accelerometer \citep{josselin1999capacitive} based on Onera's expertise in the field of accelerometry and gravimetry: CHAMP, GRACE and GOCE missions \citep{touboul1999accelerometers,marque2008ultra} and the upcoming Microscope mission \citep{guiu2007calibration,hudson2007development,touboul2001microscope}. Ready-to-fly technology is used with original developments aimed at reducing power consumption, size and weight. The accelerometer is mounted on a rotating stage, called Bias Rejection System, which allows removing the bias and therefore making absolute measurements. An Interface and Control Unit (ICU) is in charge of interfacing the instrument with the spacecraft.

The principle of MicroSTAR is to detect by capacitive measurement any motion of the proof mass with respect to the cage and to act, via a feedback control loop, on the proof mass thanks to electrostatic forces in order to keep it at the center of the cage. So the proof mass is almost motionless with respect to the cage in the control bandwidth of the control loop and at the level of the capacitive measurement noise.

\subsection{Design}

\begin{figure}[ht]
  \begin{center}
    \includegraphics[width=0.45 \linewidth]{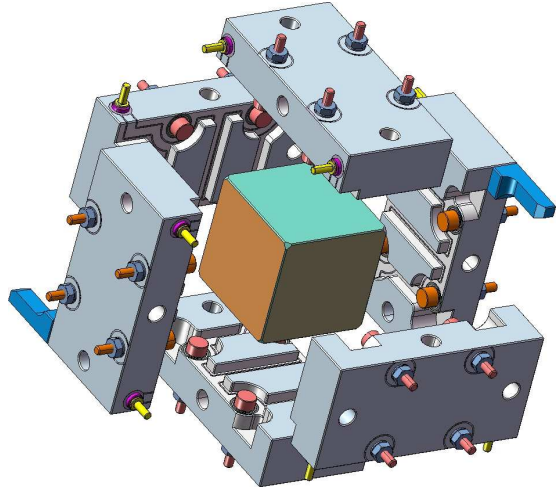}
    \caption{Exploded view of the mechanical core of MicroSTAR. The position and attitude of the proof mass is controlled by six pairs of electrodes whose potentials are defined by a control loop. The potentials $\pm V_p$ on the proof mass are kept precisely constant by means of reference voltage generators connected with two thin gold wires to the proof mass. On each ULE plates, four stops prevent the proof mass from touching the electrodes.}
    \label{fig:MicroSTAR_core}
  \end{center}
\end{figure}

The proof mass is inside a cage composed of six identical plates. This choice has been made to decrease the cost and production time of MicroSTAR (Fig. \ref{fig:MicroSTAR_core}). All these components are made in the same Ultra-Low Expansion (ULE) silica glass manufactured by Corning. The proof mass is a cube whose edge measures 2 cm and which weights 18 g. It is gold coated with a cutting such that two opposite faces have opposite polarisation. On each ULE plates, two electrodes are engraved by Ultra-Sonic machining, which is patented by Onera \citep{USpatent4934103}. The position and attitude of the proof mass with respect to the cage (6 degrees of freedom) are controlled by the six pairs of electrodes, such that there is no contact between the proof mass and the cage. The electrodes are designed such that degrees of freedom are coupled two by two. And the design of the accelerometer is completely symmetric. Consequently, the accelerometer cannot be levitated under 1 g: the condition and performance of the instrument will have to be tested in a drop tower \citep{dittus1991drop}, as it is foreseen for Microscope accelerometers \citep{touboul2001microscope}.

\begin{figure}[ht]
  \begin{center}
    \includegraphics[width=0.8 \linewidth]{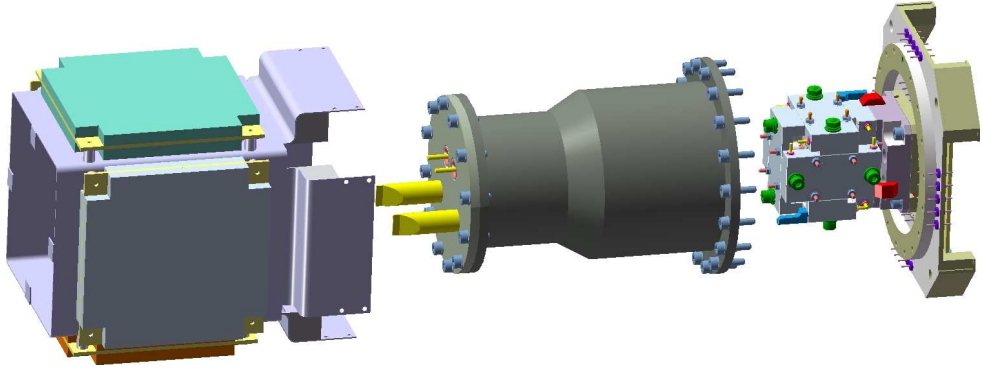}
    \caption{The core of MicroSTAR on the sole-plate with the hermetic housing. The proximity electronics is around the housing.}
    \label{fig:MicroSTAR_housing}
  \end{center}
\end{figure}

The core of MicroSTAR is mounted on an hermetic sole-plate, and an hermetic housing ensures vacuum and cleanliness. A getter maintains a vacuum smaller than $10^{-5}$ Pa during the whole life of the instrument. The proximity electronics is mounted on a Printed Card Board (PCB) around the hermetic housing.

The mass estimate for the accelerometer shown in figure \ref{fig:MicroSTAR_housing} is $1.4$ kg. The Bias Rejection System which will be added to MicroSTAR shall not exceed 1.1 kg and the Interface and Control Unit 1.0 kg. The total mass required to host the instrument on board a spacecraft will therefore be 3.1 kg.

\subsection{Control of the proof mass}

\begin{figure}[ht]
  \begin{center}
    \includegraphics[width=0.45 \linewidth]{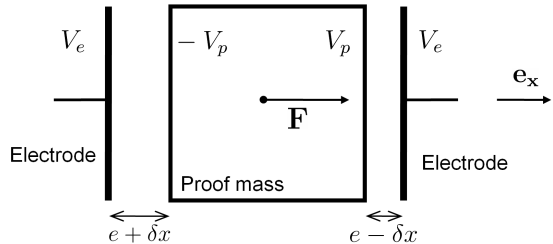}
    \caption{Schematic of a one-dimensional electrostatic accelerometer. The potentials on the proof mass $\pm V_p$ are set via two gold wires and the position of the proof mass $\delta x$ is controlled thanks to the potential on the electrodes $V_e$.}
    \label{fig:force}
  \end{center}
\end{figure}

Figure \ref{fig:force} shows in one dimension the principle of the electrostatic accelerometer. The force acting on the proof mass due to one pair of electrodes, $\ve{F}$, is equal to
\begin{equation}
  \ve{F} = \left[ \frac{\epsilon_0 \Sigma}{2(e-\delta x)^2}(V_e-V_p)^2 -  \frac{\epsilon_0 \Sigma}{2(e+\delta x)^2}(V_e+V_p)^2 \right]\ve{e_x}
\end{equation}
where $\epsilon_0 = 1/(\mu_0 c^2)$ is the electric constant \citep{mohr2008codata2}, $\Sigma$ is the surface of one electrode, $e$ is the gap between the proof mass and one electrode when the proof mass is at the center of the cage ($\delta x = 0$), $V_p$ is the potential on one side of the proof mass and $V_e$ is the potential on the electrodes. Assuming that $\delta x \ll e$, which is a valid assumption when the control loop is running, one obtains
\begin{equation}
  \frac{1}{m_A} \ve{F} = \left[ {\omega_p}^2 \delta x + G_e V_e + \frac{{\omega_p}^2}{{V_p}^2}\delta x {V_e}^2 \right] \ve{e_x}
  \label{eq:force_electro}
\end{equation}
where $m_A$ is the mass of the proof mass, ${\omega_p}^2 = 2\epsilon_0 \Sigma {V_p}^2 /(m_A e^3)$ is the electrostatic stiffness and $G_e = -2\epsilon_0 \Sigma V_p /(m_A e^2)$ is the electrostatic gain of the accelerometer. For MicroSTAR, the total force acting along one direction is the sum of the forces of the two pairs of electrodes perpendicular to this direction; and the torque acting along one direction is proportional to the difference of the forces of the two pairs of electrodes.

Note that, contrary to an harmonic oscillator, this force tends to bring the proof mass out of equilibrium because of the positive sign before $\delta x$. Therefore, a control loop is required to maintain the proof mass at the center of the cage. When $\delta x = 0$, the force acting on the proof mass is proportional to the potential of the electrodes $V_e$. Therefore, the output of the accelerometer is the quantity $G_e V_e$. The coefficient $G_e$ depends on the value of $V_p$. To keep it perfectly constant, the proof mass is linked to a reference voltage source via two gold wires, such that two opposite faces of the proof mass have an opposite potential, as in figure \ref{fig:force}.

This set up is different from previous accelerometers designed at Onera for which the potential on the proof mass was uniformly equal to $V_p$ and the potentials on the electrodes were $+V_e$ and $-V_e$. This choice has been made for MicroSTAR for miniaturization purpose as this new set up allows reducing the mass of the electronics.

\begin{table}[ht]
\caption{Parameters of MicroSTAR.}
\label{table:MicroSTAR_parameters}
  \begin{center}
    \begin{tabular}{ l  l  l }
      \hline
       & Size of the proof mass & $20\times20\times20$ mm$^3$ \\
      \hline
      $m_A$ & Mass of the proof mass & $18$ g \\
      $\Sigma$ & Size of one electrode & $94$ mm$^2$ \\
      $e$ & Gap between the proof mass and the electrode & $0.3$ mm \\
      $I_A$ & Inertia of the proof mass & $1200$ g.mm$^2$ \\
      \hline
    \end{tabular}
  \end{center}
\end{table}

\subsection{Electronics}

The electronics is composed of capacitive detectors (one for each couple of electrodes), a PID (Proportional-Integral-Derivative) control and a recombination device. As described above, the outputs of the accelerometer are the voltage applied on each of the electrodes to control the proof mass. These voltages are digitalized through an accurate Analog to Digital Converter of 24 bits and sent to the Interface and Control Unit.

The same electrodes are used for detection and action. In order to separate these two functions, the detection is done at 100 kHz by adding to the constant voltage $V_p$ a voltage $V_d$ modulated at this detection frequency. The consumption estimate for the electronics is $1.4$ W.

\subsection{Spectral characterization of MicroSTAR's noise}\label{subsection:performance}

\begin{figure}[htbp]
  \begin{center}
    \includegraphics[width=0.8 \linewidth]{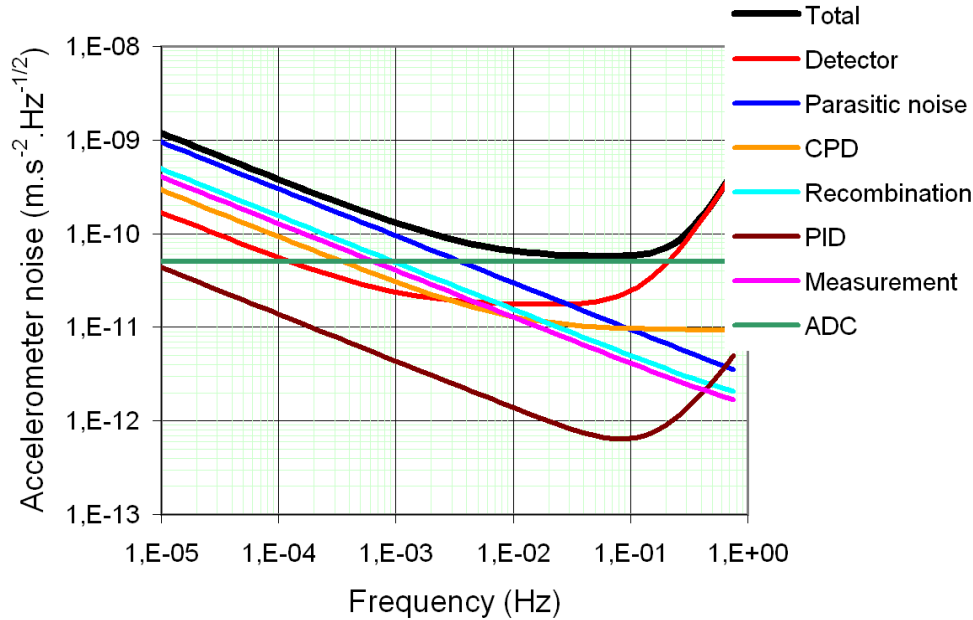}
    \caption{Square-root of the Power Spectrum Density of MicroSTAR's noise. The main contributors are the gold wires at low frequencies, the capacitive detector at high frequencies and the Analog--to--Digital Converter for frequencies between $10^{-2}$ and $10^{-1}$ Hz.}
    \label{fig:DSP}
  \end{center}
\end{figure}

In this section, the noise of the instrument is briefly described. The analysis relies on models \citep{grassia2000quantum,chhun2007equivalence,touboul2009microscope}, ground experiments \citep{willemenot1997pendule} and in-orbit measurements \citep{flury2008precise,marque2010accelerometers}. Figure \ref{fig:DSP} shows, for the parameters given in table \ref{table:MicroSTAR_parameters} and for a measurement range equal to $1.8 \times 10^{-4}$ m s$^{-2}$, the square-root of the power spectrum density of the noise and the different contributions:

\begin{itemize}
    \item The Proportional-Integral-Derivative controller (PID) is a minor contributor. This function is performed using operational amplifiers.
    \item The ``Recombination'' term corresponds to the noise induced by the coupling of two degrees of freedom. The controller computes the translation and rotation which must be applied to the proof mass. These quantities then have to be translated into forces that must be applied by the electrodes. Some noise is generated in this process by the amplifiers which are used to perform this function.
    \item The Contact-Potential-Difference (CPD) refers to voltage fluctuations on the electrodes, also called patch effect \citep{speake1996forces}. The chemical potential, which is the sum of the electrical potential and the work function, is constant on the surface. The spatial variations of the electrical potential come from the spatial variations of the work function, which is due to atomic composition of the materials. In addition, there are temporal fluctuations due to evolution of the material characteristics.
    \item The ``parasitic noise'' term includes the several contributions: radiometer effect, asymmetric outgassing, piston noise, thermal radiation pressure noise and the two gold wires, which are the main contributors.
    \item The ``Detector'' term corresponds to the noise generated by the electronics in charge of measuring the proof mass position via the measurement of the capacity of each gap. It is the main contributor at high frequencies.
    \item The ``Measurement'' term corresponds to the noise of the amplifiers used to measure the potential on the electrodes.
    \item The Analog-to-Digital Converter (ADC) introduces a quantification noise in the measurement chain which is constant for all frequencies. Its value increases with the range of the instrument and decreases with the number of bits available.
\end{itemize}

An analytic formula of the square root of MicroSTAR noise power spectrum density as a function of frequency can be given for simulation purpose:
\begin{equation}
  \sqrt{S_{n}(f)} = K \sqrt{ 1 + \left(\frac{f}{4.2 \ \mathrm{mHz}}\right)^{-1} + \left(\frac{f}{0.27 \ \mathrm{Hz}}\right)^{4} }
  \label{eq:acc_noise}
\end{equation}
with
\begin{equation}
  K = 5.7 \times 10^{-11} \ \mathrm{m}.\mathrm{s}^{-2}.\mathrm{Hz}^{-1/2}
\end{equation}

\section{Measurement principle}\label{section:measurement}

This section describes how GAP is used to perform bias-free measurements. First, the link between the non-gravitational acceleration and the electrostatic force (cf. equation (\ref{eq:force_electro})) is established. The bias of the instruments appears in the relation derived and it is explained how to use the Bias Rejection System to remove it.

\subsection{Electrostatic measurement of the non-gravitational accelerations}

The link between the force described by equation (\ref{eq:force_electro}) and the non-gravitational acceleration acting on the spacecraft is established in this section. Let us assume that $R_0\left(O,\ve{e^0_x},\ve{e^0_y},\ve{e^0_z}\right)$ is the reference frame of the Solar System, which is supposed to be Galilean. $R_1\left(S,\ve{e^1_x},\ve{e^1_y},\ve{e^1_z}\right)$ is the reference frame attached to the satellite and $R_2\left(C,\ve{e^2_x},\ve{e^2_y},\ve{e^2_z}\right)$ is the reference frame attached to MicroSTAR. $O$ is the barycenter of the Solar System, $C$ the center of the cage of MicroSTAR, and $S$ the inertia center of the spacecraft. The angular velocity of the frame $R_i$ with respect to the frame $R_j$ is called $\ve{\Omega_{i/j}}$. The Bias Rejection System allows MicroSTAR to rotate with respect to the satellite: it will be assumed that $R_2$ is rotating around $\ve{e^1_z}$ with respect to $R_1$ and that the rotation is parametrized by the angle $\theta$. This means that $\ve{\Omega_{2/1}} = \dot{\theta}\ve{e^1_z}$.

Below is a list of the forces acting on the proof mass and on the spacecraft. $S$ and $A$ represent respectively the satellite and the proof mass. A difference is made between the gravitational forces and the non-gravitational forces:
\begin{itemize}
    \item $\ve{F^{NG}_{\mathnormal{ext}\rightarrow \mathnormal{S}}}$: Non-gravitational force acting on the satellite due to the external environment. It includes forces discussed in section \ref{subsection:NGacc}.
    \item $\ve{F^{G}_{\mathnormal{ext} \rightarrow \mathnormal{S}}}$: Gravitational force acting on the satellite due to the external environment. This force is induced by the mass distribution in the Solar System.
    \item $\ve{F^{NG}_{\mathnormal{ext} \rightarrow \mathnormal{A}}}$: Non-gravitational force acting on the proof mass due to the external environment.
    \item $\ve{F^{G}_{\mathnormal{ext} \rightarrow \mathnormal{A}}}$: Gravitational force acting on the proof mass due to the external environment.
    \item $\ve{F^{G}_{\mathnormal{S} \rightarrow \mathnormal{A}}}$: Gravitational force acting on the proof mass due to the satellite. In the following, this force will be referred to as self-gravity.
    \item $\ve{F^{gold}_{\mathnormal{S} \rightarrow \mathnormal{A}}}$: Force acting on the proof mass due to the satellite via the gold wires.
    \item $\ve{F^{elec}_{\mathnormal{S} \rightarrow \mathnormal{A}}}$: Force acting on the proof mass due to the satellite via the six  pairs of electrodes (cf. equation (\ref{eq:force_electro})).
\end{itemize}

By applying Newton's second law to the satellite and to the proof mass, one obtains the following equation:
\begin{equation}
  \left( \frac{1}{m_A} + \frac{1}{m_S} \right) \ve{F^{elec}_{\mathrm{S}\rightarrow \mathrm{A}}}  =  \frac{1}{m_S}\ve{F^{NG}_{\mathnormal{ext} \rightarrow \mathnormal{S}}} + \ve{A_1} - \ve{A_2} - \ve{A_3}
  \label{eq:dynamics_final}
\end{equation}
where it is assumed that the proof mass remains at the center of the cage due to the control loop -- i.e. $R_1$ and $R_2$ are identical -- and the satellite is perfectly rigid -- i.e. the distance between $C$ and $S$ remains constant. $m_S$ and $m_A$ are respectively the masses of the satellite and of the proof mass. $\ve{A_1}$, $\ve{A_2}$ and $\ve{A_3}$ are equal to:
\begin{equation}
  \ve{A_1} = \ve{\dot{\Omega}_{1/0}} \wedge\ve{SC} + \ve{\Omega_{1/0}} \wedge \left(\ve{\Omega_{1/0}} \wedge\ve{SC} \right)
\end{equation}
\begin{equation}
  \ve{A_2} =  \left( \frac{1}{m_A} + \frac{1}{m_S} \right) \ve{F^{G}_{\mathnormal{S}\rightarrow \mathnormal{A}}} + \frac{1}{m_A} \ve{F^{NG}_{\mathnormal{ext} \rightarrow \mathnormal{A}}} + \left( \frac{1}{m_A} \ve{F^{G}_{\mathnormal{ext}\rightarrow \mathnormal{A}}} - \frac{1}{m_S} \ve{F^{G}_{\mathnormal{ext}\rightarrow \mathnormal{S}}}  \right)
\end{equation}
\begin{equation}
  \ve{A_3} =  \left( \frac{1}{m_\mathnormal{A}} + \frac{1}{m_\mathnormal{S}} \right) \ve{F^{gold}_{\mathnormal{S}\rightarrow \mathnormal{A}}}
\end{equation}

In the next section, the two terms $\ve{A_1}$ and $\ve{A_2}$ will not be taken into account because they are either negligible or known with sufficient accuracy to be removed. This hypothesis will be discussed in section \ref{section:accommodation}. Moreover, to simplify the discussion, it will be assumed that $\ve{\Omega_{1/0}} = \ve{0}$, i.e. the spacecraft is not rotating with respect to the reference frame of the Solar System.

\subsection{Bias rejection}\label{section:bias_rejection}

Let's introduce $F^{elec}_{2,x}$, $F^{elec}_{2,y}$ and $F^{elec}_{2,z}$, the components of $\ve{F^{elec}_{\mathnormal{S}\rightarrow \mathnormal{A}}}$ in $R_2$; $F^{gold}_{2,x}$, $F^{gold}_{2,y}$ and $F^{gold}_{2,z}$, the components of $\ve{F^{gold}_{\mathnormal{S}\rightarrow \mathnormal{A}}}$ in $R_2$; and $F^{NG}_{0,x}$, $F^{NG}_{0,y}$ and $F^{NG}_{0,z}$, the components of $\ve{F^{NG}_{ext\rightarrow \mathnormal{S}}}$ in $R_0$.

The electronics of the instruments gives the values of the components of $\ve{F^{elec}_{\mathnormal{S}\rightarrow \mathnormal{A}}}$ in $R_2$. However, there are defaults in the measurement chain (cf. equation (\ref{eq:bias_noise}) with $\delta k_1=k_2=0$) and this translates in particular in a bias called ``electronic bias''. If $m_x$, $m_y$ and $m_z$ are the outputs of MicroSTAR for each axis, the following relations hold
\begin{equation}
  m_\zeta = F^{elec}_{2,\zeta} + b_\zeta + n_\zeta, \ \ \zeta\in\{x,y,z\}
\end{equation}
where $b_\zeta$ is the electronic bias and $n_\zeta$ is the measurement noise. Moreover, equation (\ref{eq:dynamics_final}) yields
\begin{eqnarray}
  F^{elec}_{2,x} & = & - F^{gold}_{2,x} + \mu \left( F^{NG}_{0,x} \cos(\theta) + F^{NG}_{0,y} \sin(\theta) \right) \\
  F^{elec}_{2,y} & = & - F^{gold}_{2,y} + \mu \left( - F^{NG}_{0,x} \sin(\theta) + F^{NG}_{0,y} \cos(\theta) \right) \\
  F^{elec}_{2,z} & = & - F^{gold}_{2,z} + \mu F^{NG}_{0,z}
\end{eqnarray}
where $\mu = m_A/(m_A + m_S) = m_A/m_S + o(m_A/m_S)$ and $\theta$ is the angle controlled by the Bias Rejection System. Considering only the plane perpendicular to the rotation axis, the previous equations lead to
\begin{eqnarray}
  m_x & = & \mu \left( F^{NG}_{0,x} \cos(\theta) + F^{NG}_{0,y} \sin(\theta) \right) + \tilde{b}_x + n_x \label{eq:mx} \\
  m_y & = & \mu  \left( - F^{NG}_{0,x} \sin(\theta) + F^{NG}_{0,y} \cos(\theta) \right) + \tilde{b}_y + n_y \label{eq:my}
\end{eqnarray}
where $\tilde{b}_\nu = b_\nu -F^{gold}_{2,\nu}$ ($\nu \in \{x;y\}$) are the total measurement bias. The non-gravitational acceleration acting on the spacecraft is modulated by the angle $\theta$ whereas the bias of MicroSTAR is not. It is therefore possible to separate the bias $\tilde{b}_x$ and $\tilde{b}_y$ from the true acceleration acting on the spacecraft, using the Bias Rejection System.

One possible way to use these two equations is to perform a spectral separation of the acceleration and of the bias. Assuming for example that $\theta = \omega t$, with $\omega$ a constant angular speed, $m_x$ and $m_y$ can be multiplied by $\cos(\omega t)$, after being received on ground. For the $x$ axis, this would lead to the demodulated quantity
\begin{equation}
\tilde{m}_x = \frac{\mu}{2} F^{NG}_{0,x} + \frac{\mu}{2} \left( - F^{NG}_{0,x} \cos(2\omega t) + F^{NG}_{0,y} \sin(2\omega t) \right) - (b_x+F^{gold}_{2,x}) \cos(\omega t) - n_x\cos(\omega t)
\end{equation}
The external acceleration which was modulated at the frequency $\omega/(2\pi)$ in $m_x$ and $m_y$ (cf. equations (\ref{eq:mx}) and (\ref{eq:my})) is brought back in the continuous by the demodulation. The bias $b_x+F^{gold}_{2,x}$ is supposed to vary such that the Fourier transform of the term $(b_x+F^{gold}_{2,x}) \cos(\omega t)$ is null at low frequencies. By applying a low-pass filter to $\tilde{m}_x$ and $\tilde{m}_y$, one recovers an unbiased measurement of $F^{NG}_{0,x}$ and $F^{NG}_{0,y}$ plus the low frequencies of the modulated noise $n_\zeta \cos(\omega t)$. Therefore, the noise is also filtered such that only the components around the modulated frequency affect the final measurements. This heterodyne method is the simplest way of achieving the recovery of the non-gravitational acceleration. It has however the disadvantage of requiring the rotating stage to operate continuously. This increases power consumption and may lead to a quicker breakdown of the components. It also requires to have a counter-rotating body in order not to transfer angular momentum to the spacecraft. Finally, the rotation may induce additional forces on the proof mass which would spoil the signal.

\begin{figure}[htbp]
  \begin{center}
    \includegraphics[width=0.45 \linewidth]{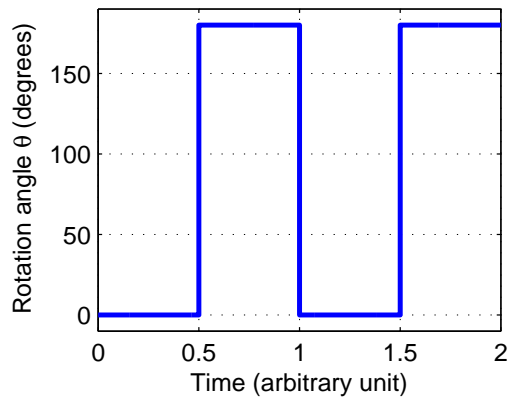}
    \caption{Modulation signal used to separate the non-gravitational acceleration from the bias of the electrostatic accelerometer. This signal is idealized since the flips from one position to the other are not instantaneous.}
    \label{fig:modulation_signal}
  \end{center}
\end{figure}

A more convenient modulation technique consists in flipping MicroSTAR by $\pi$ rad, leading to a signal with a period called $\tau$. This modulation scheme, shown in figure \ref{fig:modulation_signal}, does not have the drawbacks of the previous one. In practice however, the rotation between 0 and $\pi$ rad is not as sharp as shown on the figure. Indeed, the angular acceleration and angular velocity are kept small enough such that there is no loss of control of the proof mass due to forces exceeding the maximum range of control of the electrostatic actuation. As it was mentioned in the previous paragraph, the data acquired when the accelerometer is rotating may be spoiled by additional forces. Therefore, in the case of the signal considered here, the data acquired when the accelerometer is rotating are not used.

Concerning the precision of the measured acceleration, called $\delta a$, it can be approximated by the following formula, whose squared value corresponds to the value of the noise spectrum at the modulated frequency divided by the integration time.
\begin{equation}
  \delta a \sim \sqrt{ \frac{1}{T} S_n\left(\frac{1}{\tau}\right) }
  \label{eq:precision}
\end{equation}
where $S_{n}(f)$ is given by equation (\ref{eq:acc_noise}), $T$ is the integration time and $\tau$ is the period of the modulation signal. This gives only an order of magnitude since the harmonics of the modulation signal are not taken into account. An exact formula for $\delta a$ can be derived but is beyond the scope of this article. Assuming that $\tau = 10$ min, that the integration time is $5$ h and that the only sources of noise and errors are the ones described in section \ref{subsection:performance}, the precision achieved on the acceleration is then equal to $0.8$ pm s$^{-2}$ according to equation (\ref{eq:precision}). The exact derivation of the uncertainty for the signal of figure \ref{fig:modulation_signal} gives $1$ pm s$^{-2}$, with a modulation period $\tau = 10$ min, a integration time $T=5$ h and a rotating time between the position 0 rad and $\pi$ rad equal to 120 s.

It is important to notice that the Bias Rejection System allows rejecting the bias of the two axes which are perpendicular to the axis of rotation. To remove the bias on the third axis, another rotating stage would be needed. The direction along which the absolute measurement of the non-gravitational acceleration is not required must therefore be the direction of the rotation axis. In the case of an interplanetary probe, this axis should be perpendicular to the orbit plane.

\subsection{Requirements on the Bias Rejection System}\label{subsection:other}

The Bias Rejection System is similar, in its principle, to filter wheel used for space camera or to rotary actuator used for solar array drives, optical mechanism drives, deployment mechanisms and antenna pointing mechanisms. This section gives the requirements this subsystem has to meet in order to obtain the target accuracy of 1 pm s$^{-2}$.

The rotating stage is controlled by the ICU through a closed loop with an encoder for detection of the rotation. The accuracy of the pointing knowledge of the rotating stage, called $\theta^*$, is crucial to obtain the target accuracy called $a^*$. Calling the external acceleration $a_{ext}$, the value of $\theta^*$ is given by
\begin{equation}
  \theta^* = \arcsin\left( \frac{a^*}{a_{ext}} \right) \approx \frac{a^*}{a_{ext}}
\end{equation}
Assuming an external acceleration equal to $10^{-7}$ m s$^{-2}$ and a target accuracy of $1$ pm s$^{-2}$, the requirement on the angle is $\theta^* = 10^{-5}$ rad.

Concerning the wobble of the rotating stage, it has to be smaller than a value called $\phi^*$ given by:
\begin{equation}
  \phi^* = \sqrt{\frac{2a^*}{a_{ext}}} \approx \sqrt{2 \theta^*}
\end{equation}
With the previous numerical values, the requirement on the wobble is $\phi^*=4.5\times 10^{-3}$ rad. The square root in the value of $\phi^*$ comes from the fact that only the term at the order ${\phi^*}^2$ induces an error on the acceleration measured by the instrument (the term of order one appears as a bias and is therefore removed).

\section{Accommodation of the instrument}\label{section:accommodation}

In the previous section, a precision of $1$ pm s$^{-2}$ on the acceleration measured by GAP was obtained for an integration time of 5 h. Equation (\ref{eq:dynamics_final}) shows however that the force measured $\ve{F^{elec}_{\mathnormal{S}\rightarrow \mathnormal{A}}}$ is proportional to the non-gravitational forces acting on the spacecraft $\ve{F^{NG}_{\mathnormal{ext}\rightarrow \mathnormal{S}}}$ plus a number of terms, $\ve{A_1}$, $\ve{A_2}$ and $\ve{A_3}$, which can induce uncertainties on the measurement. $\ve{A_3}$, which describes the gold wire action was taken into account, and the two other terms, $\ve{A_1}$ and $\ve{A_2}$, were supposed to be negligible. This section discusses this hypothesis. $\ve{A_1}$ describes that the spacecraft is not a Galilean reference frame; and the components of $\ve{A_2}$ are approximately constant in $R_0$ which means that it can be considered as an external acceleration which is modulated.

As soon as the instrument is not at the center of inertia of the spacecraft and the spacecraft is rotating with respect to the Solar System reference frame, term $\ve{A_1}$ is not null. It has not been taken into account because it was supposed to be known  precisely enough. This knowledge requires to measure with an accuracy compatible with the target accuracy of the instrument, the attitude of the satellite, its angular velocity and the position of the proof mass with respect to the center of mass of the satellite, i.e. the vector $\ve{SC}$. The attitude of the satellite and its angular velocity are measured with star trackers. The knowledge of $\ve{SC}$ requires to measure the position of the instrument with respect to the satellite prior to launch and to precisely monitor the position of the center of inertia of the spacecraft, which may change due to the use of propellant. This knowledge, along with the mass of the spacecraft, is also needed to estimate the self-gravity $\ve{F^{G}_{\mathnormal{S}\rightarrow \mathnormal{A}}}$ which appears in the term $\ve{A_2}$.

The only non-gravitational force, $\ve{F^{NG}_{\mathnormal{ext}\rightarrow\mathnormal{A}}}$, possibly acting on the proof mass due to the exterior environment is electro-magnetic because the proof mass is protected by a hermetic housing. Depending on the spacecraft on board which the instrument will be, this force may not be negligible. Therefore, it may be required to measure the magnetic field in order to remove this contribution from the measurements.

The difference of the gravitational acceleration can be constrained. By introducing $r_s$ the distance from the Sun to the inertia center of the spacecraft and $\delta r$ the distance between $S$ and $A$, the following upper bound comes:
\begin{equation}
  \left|\left| \frac{1}{m_A} \ve{F^{G}_{\mathnormal{ext}\rightarrow\mathnormal{A}}} - \frac{1}{m_S} \ve{F^{G}_{\mathnormal{ext}\rightarrow\mathnormal{S}}} \right|\right| \leq \frac{2GM_\odot}{{r_s}^3} \delta r
\end{equation}
where $G$ is the gravitational constant and $M_\odot$ the mass of the Sun. Given conservative assumptions -- the spacecraft is at one astronomical unit from the Sun and $\delta r$ is equals to 1 meter -- the upper bound is equal to $7.9\times 10^{-14}$ m s$^{-2}$, which is one order of magnitude smaller than $1$ pm s$^{-2}$.

This short description of the ``parasitic'' terms in the measurement shows that the accommodation of the instrument in a spacecraft requires care. Self-gravity may be reduced and better controlled if the instrument is placed at the center of mass of the spacecraft. As for the pointing system and the attitude control of the spacecraft, it must be compatible with the accuracy target of the instrument.

\section{Conclusion}\label{section:conclusion}

The electrostatic accelerometer with bias rejection GAP, designed to answer the need for large scale tests of General Relativity and for planetary observations, relies on a new concept to make measurements of the non-gravitational acceleration with no bias. The main idea is to mount the electrostatic accelerometer MicroSTAR on a rotating stage, so that the external signal to be measured is modulated and distinguished from the bias of the accelerometer after post-treatment.

The article presented the design of MicroSTAR as well as its spectral characterization. It also gave the main specifications on the Bias Rejection System. For a given modulation scheme, it is possible to derive the final precision on the acceleration measurement: with an integration time of five hours and a modulation period of 10 min, the precision is $1$ pm s$^{-2}$ on the measured acceleration.

This opens the possibility to improve the accuracy on the comparison between the motion of the spacecraft and theoretical predictions by three orders of magnitude with respect to existing techniques. This accuracy leads to new applications for large scale tests of General Relativity as well as for improved planetary observations.

\section{Acknowledgments}\label{section:acknowledgments}
The authors are grateful to CNES for its financial support and to the members of the Odyssey team and of the GAP team for their collaboration: J.-M. Courty (LKB, France), H. Dittus, T. van Zoest (Institute of Space Systems, DLR, Germany), C. L\"ammerzahl, H. Selig (ZARM, Germany), E. Hinglais, S. L\'eon-Hirtz (CNES, France), G. M\'etris (OCA, France), F. Sohl (Institute of Planetary Research, DLR, Germany), P. Touboul (Onera, France) and P. Wolf (SYRTE, France).





\begin{thebibliography}{}

\bibitem[Aguirre et al.(2001)]{aguirre2001astrophysical}
Aguirre, A., Burgess, C.P., Friedland, A., \& Nolte, D.,
Astrophysical constraints on modifying gravity at large distances, Class. Quantum Grav., 18, R223, 2001.

\bibitem[Anderson et al.(1998)]{anderson1998indication}
Anderson, J.D., Laing, P.A., Lau, E.L., Liu, A.S., Nieto, M.M., \& Turyshev, S.G.,
Indication, from Pioneer 10/11, Galileo, and Ulysses Data, of an Apparent Anomalous, Weak, Long-Range Acceleration, Phys. Rev. D 81, 2858, 1998.

\bibitem[Anderson et al.(2002a)]{anderson2002study}
Anderson, J.D., Laing, P.A., Lau, E.L., Liu, A.S., Nieto, M.M., \& Turyshev, S.G.,
Study of the anomalous acceleration of Pioneer 10 and 11, Phys. Rev. D 65, 082004, 2002.

\bibitem[Anderson et al.(2002b)]{anderson2002mission}
Anderson, J.D., Nieto, M.M., \& Turyshev, S.G.,
A Mission to Test the Pioneer Anomaly, Int. J. Mod. Phys. D 11, 1545-1552, 2002.

\bibitem[Anderson \& Nieto(2009)]{anderson2009astrometric}
Anderson, J.D., \& Nieto, M.M.,
Astrometric solar-system anomalies, Proc. Int. Astron. Union 5, 189-197, 2009.

\bibitem[Asmar et al.(2005)]{asmar2005spacecraft}
Asmar, S.W., Armstrong, J.W., Iess, L., \& Tortora, P.,
Spacecraft Doppler tracking: Noise budget and accuracy achievable in precision radio science observations, Radio Sci. 40, RS2001, 2005.

\bibitem[Bertolami \& Paramos(2007)]{bertolami2007mission}
Bertolami, O., \& P\'aramos, J.,
A mission to test the Pioneer anomaly: estimating the main systematic effects, Int. J. Mod. Phys. D16, 1611-1623, 2007.

\bibitem[Bertolami et al.(2008)]{bertolami2008thermal}
Bertolami, O., Francisco, F., Gil, P.J.S., \& P\'aramos, J.,
Thermal analysis of the Pioneer anomaly: A method to estimate radiative momentum transfer, Physical Review D 78, 103001, 2008.

\bibitem[Biesbroek(2008)]{biesbroek2008laplace}
Biesbroek, R.,
Laplace: Assessment of the Jupiter Ganymede Orbiter, CDF Study Report CDF-77(A), ESA, 2008.

\bibitem[Blanc et al.(2005)]{blanc2005tracing}
Blanc, M., Moura, D., Alibert, Y., et al.,
Tracing the origins of the Solar System, in Trends in Space Science and Cosmic Vision 2020, ESA Special Publications 588, 213-224, 2005.

\bibitem[Bruneton \& Esposito-Farese(2007)]{bruneton2007field}
Bruneton, J.P., \& Esposito-Farese, G.,
Field-theoretical formulations of MOND-like gravity, Phys. Rev. D 76, 124012, 2007.

\bibitem[Campergue et al.(1990)]{USpatent4934103}
Campergue, G., Gouhier, R., Horriere, D., \& Thiriot, A.,
Machine for ultrasonic abrasion machining, US patent 4934103, 1990.

\bibitem[Chhun et al.(2007)]{chhun2007equivalence}
Chhun, R., Hudson, D., Flinoise, P., Rodrigues, M., Touboul, \& P., Foulon, B.,
Equivalence principle test with Microscope: Laboratory and engineering models preliminary results for evaluation of performance, Acta Astronaut. 60, 873-879, 2007.

\bibitem[Christophe et al.(2008)]{christophe2008gravity}
Christophe, B., Foulon, B., L\'evy, A., et al.,
Gravity Advanced Package, an accelerometer package for Laplace or Tandem missions, in: Charbonnel, C., Combes, F., Samadi, R. (Eds.), SF2A 2008, Soci\'et\'e Fran\c caise d'Astronomie et d'Astrophysique, p. 103, 2008.

\bibitem[Christophe et al.(2009)]{christophe2009odyssey}
Christophe, B., Andersen, P.H., Anderson, J.D., et al.,
Odyssey: a Solar System Mission, Exp. Astron. 23, 529-547, 2009.

\bibitem[Copeland et al.(2006)]{copeland2006dynamics}
Copeland, E.J., Sami, M.,\& Tsujikawa, S.,
Dynamics of dark energy, Int. J. Mod. Phys. D 15, 1753-1935, 2006.

\bibitem[Dittus(1991)]{dittus1991drop}
Dittus, H.,
Drop tower 'bremen': A weightlessness laboratory on earth, Endeavour 15, 72-78, 1991.

\bibitem[Dittus et al.(2005)]{dittus2005mission}
Dittus, H., Turyshev, S.G., L\"ammerzahl, C., et al.,
A Mission to Explore the Pioneer Anomaly, in Trends in Space Science and Cosmic Vision 2020, ESA Special Publications vol.588, 3-10, 2005.

\bibitem[ESA(2010)]{esa2010roadmap}
ESA (Fundamental Physics Roadmap Advisory Team),
A Roadmap for Fundamental Physics in Space, 2010. Available at [08/23/2010]: \url{http://sci.esa.int/fprat}

\bibitem[Fienga et al.(2010)]{fienga2010gravity}
Fienga, A., Laskar, J., Kuchynka, P., Leponcin-Lafitte, C., Manche, H., \& Gastineau, M.,
Gravity tests with INPOP planetary ephemerides, Proceedings IAU Symposium 261, 159-169, 2010.

\bibitem[Flury et al.(2008)]{flury2008precise}
Flury, J., Bettadpur, S., \& Tapley, B.,
Precise accelerometry onboard the GRACE gravity field satellite mission, Adv. Space Res. 42, 1414-1423, 2008.

\bibitem[Frieman(2008)]{frieman2008dark}
Frieman, J. A., Turner, M. S., \& Huterer, D.,
Dark Energy and the Accelerating Universe, Annual Review of Astronomy and Astrophysics 46, 385, 2008.

\bibitem[Fr\"ohlich \& Lean(2004)]{frohlich2004solar}
Fr\"ohlich, C., \& Lean, J.,
Solar radiative output and its variability: evidence and mechanisms, Astron. Astrophys. Rev. 12, 273-320, 2004.

\bibitem[Grassia et al.(2000)]{grassia2000quantum}
Grassia, F., Courty, J.M., Reynaud, S., \& Touboul, P.,
Quantum theory of fluctuations in a cold damped accelerometer, Eur. Phys. J. D 8, 101-110, 2000.

\bibitem[Guiu et al.(2007)]{guiu2007calibration}
Guiu, E., Rodrigues, M., Touboul, P., \& Pradels, G.,
Calibration of MICROSCOPE, Adv. Space Res. 39, 315-323, 2007.

\bibitem[Hudson et al.(2007)]{hudson2007development}
Hudson, D., Chhun, R., \& Touboul, P.,
Development status of the differential accelerometer for the MICROSCOPE mission, Adv. Space Res. 39, 307-314, 2007.

\bibitem[Jaekel \& Reynaud(2005a)]{jaekel2005gravity}
Jaekel, M.T., \& Reynaud, S.,
Gravity tests in the solar system and the Pioneer anomaly, Mod. Phys. Lett. A 20, 1047-1055, 2005.

\bibitem[Jaekel \& Reynaud(2005b)]{jaekel2005post}
Jaekel, M.T., \& Reynaud, S.,
Post-Einsteinian tests of linearized gravitation, Class. Quantum Grav. 22, 2135-2158, 2005.

\bibitem[Jaekel \& Reynaud(2006a)]{jaekel2006post}
Jaekel, M.T., \& Reynaud, S.,
Post-Einsteinian tests of gravitation, Class. Quantum Grav. 23, 777-798, 2006.

\bibitem[Jaekel \& Reynaud(2006b)]{jaekel2006radar}
Jaekel, M.T., \& Reynaud, S.,
Radar ranging and Doppler tracking in post-Einsteinian metric theories of gravity, Class. Quantum Grav. 23, 7561-7579, 2006.

\bibitem[Johann et al.(2008)]{johann2008exploring}
Johann, U., Dittus, H., \& L\"ammerzahl, C.,
Exploring the Pioneer Anomaly, in Lasers, Clocks and Drag-Free Control, eds H. Dittus et al., 577-604, 2008.

\bibitem[Josselin et al.(1999)]{josselin1999capacitive}
Josselin, V., Touboul, P., \& Kielbasa, R.,
Capacitive detection scheme for space accelerometers applications, Sens. Actuators A 78, 92-98, 1999.

\bibitem[L\"ammerzahl et al.(2008)]{lammerzahl2008physics}
L\"ammerzahl, C., Preuss, O., Dittus, H.,
Is the Physics Within the Solar System Really Understood?,
Lasers, Clocks and Drag-Free Control,
Astrophysics and Space Science Library 349, 2008.

\bibitem[L\'evy et al.(2009)]{levy2009pioneer}
L\'evy, A., Christophe, B., B\'erio, P., M\'etris, G., Courty, J.M., \& Reynaud, S.,
Pioneer 10 Doppler data analysis: disentangling periodic and secular anomalies, Adv. Space Res. 43, 1538-1544, 2009.

\bibitem[Markwardt(2002)]{markwardt2002independent}
Markwardt, C.,
Independent Confirmation of the Pioneer 10 Anomalous Acceleration, 2002. arXiv:gr-qc/0208046v1

\bibitem[Marque et al.(2008)]{marque2008ultra}
Marque, J.P., Christophe, B., Liorzou, F., Bodovill\'e, G., Foulon, B., Gu\'erard, J., \& Lebat, V.,
The Ultra Sensitive Accelerometers of the ESA GOCE Mission, 59th International Astronautical Congress, 2008.

\bibitem[Marque et al.(2010)]{marque2010accelerometers}
Marque, J.P., Christophe, \& Foulon, B.,
Accelerometers of the GOCE mission: return of experience from one year of in-orbit, Living Planet Symposium, 2010.

\bibitem[Milani et al.(2001)]{milani2001gravity}
Milani, A., Rossi, A., Vokrouhlick\'y, D., Villani, D. \& Bonanno, C.,
Gravity field and rotation state of Mercury from the BepiColombo Radio Science Experiments. Planet. Space Sci. 49, 1579-1596, 2001.

\bibitem[Moffat(2005)]{moffat2005gravitational}
Moffat, J.W.,
Gravitational theory, galaxy rotation curves and cosmology without dark matter, J. Cosmol. Astropart. Phys. 2005, 003, 2005.

\bibitem[Moffat(2006)]{moffat2006scalar}
Moffat, J.W.,
Scalar-tensor-vector gravity theory, J. Cosmol. Astropart. Phys. 2006, 004, 2006.

\bibitem[Mohr et al.(2008)]{mohr2008codata2}
Mohr, P.J., Taylor, B.N., \& Newell, D.B.,
CODATA recommended values of the fundamental physical constants: 2006, Rev. Mod. Phys. 80, 633-730, 2008.

\bibitem[Moyer(2008)]{moyer2000formulation}
Moyer, T.D.,
Formulation for Observed and Computed Values of Deep Space Network Data Types for Navigation, Deep Space Communications and Navigation Series 2, 2000.

\bibitem[Nieto et al.(2005)]{nieto2005directly}
Nieto, M.M., Turyshev, S.G. \& Anderson, J.D.,
Directly measured limit on the interplanetary matter density from Pioneer 10 and 11, Phys. Lett. B 613, 11-19, 2005.

\bibitem[Nojiri \& Odintsov(2007)]{nojiri2007introduction}
Nojiri, S., \& Odintsov, S.D.,
Introduction to modified gravity and gravitational alternative for dark energy, Int. J. Geom. Meth. Mod. Phys. 4, 115-145, 2007.

\bibitem[Olsen(2007)]{olsen2007constancy}
Olsen, O.,
The constancy of the Pioneer anomalous acceleration, Astron. Astrophys. 463, 393-397, 2007.

\bibitem[Reynaud \& Jaekel(2008)]{reynaud2008tests}
Reynaud, S., \& Jaekel, M.T.,
Tests of general relativity in the solar system, in Atom Optics and Space Physics, eds E. Arimondo et al., 203-217 (IOS Press, 2009).

\bibitem[Reynaud et al.(2009)]{reynaud2009testing}
Reynaud, S., Salomon, C., \& Wolf, P.,
Testing General Relativity with Atomic Clocks, Space Sci. Rev. 148, 233-247, 2009.

\bibitem[Rievers et al.(2010)]{rievers2010modeling}
Rievers, B., L\"ammerzahl, C., \& Dittus, H.,
Modeling of Thermal Perturbations Using Raytracing Method with Preliminary Results for a Test Case Model of the Pioneer 10/11 Radioisotopic Thermal Generators, Space Sci.Rev. 151, 123-133, 2010.

\bibitem[Speake(1996)]{speake1996forces}
Speake, C.C.,
Forces and force gradients due to patch fields and contact-potential differences, Class. Quantum Grav. 13, A291, 1996.

\bibitem[Tortora et al.(2004)]{tortora2004precise}
Tortora, P., Iess, L., Bordi, J.J., Ekelund, J.E., \& Roth, D.C.,
Precise Cassini navigation during solar conjunctions through multifrequency plasma calibrations, Journ. Guidance Control Dyn. 27, 251-257, 2004.

\bibitem[Toth \& Turyshev(2009)]{toth2009thermal}
Toth, V.T., \& Turyshev, S.G.,
Thermal recoil force, telemetry, and the Pioneer anomaly, Phys. Rev. D 79, 43011, 2009.

\bibitem[Touboul et al.(1999)]{touboul1999accelerometers}
Touboul, P., Willemenot, E., Foulon, B., \& Josselin, V.,
Accelerometers for CHAMP, GRACE and GOCE space missions: synergy and evolution, Boll. Geof. Teor. Appl. 40, 321-327, 1999.

\bibitem[Touboul \& Rodrigues(2001)]{touboul2001microscope}
Touboul, P., \& Rodrigues, M.,
The MICROSCOPE space mission, Class. Quantum Grav. 18, 2487-2498, 2001.

\bibitem[Touboul(2009)]{touboul2009microscope}
Touboul, P.,
The Microscope Mission and Its Uncertainty Analysis, Space Sci. Rev. 148, 455-474, 2009.

\bibitem[Turyshev \& Toth(2009)]{turyshev2009pioneer}
Turyshev, S.G., \& Toth, V.T.,
The Pioneer Anomaly in the Light of New Data, Space Sci. Rev. 148, 149-167, 2009.

\bibitem[Turyshev \& Toth(2010)]{turyshev2010pioneer}
Turyshev, S.G., \& Toth, V.T.,
The Pioneer anomaly, Living Reviews in Relativity 13, 4, 2010. Available at [2010/10/13]: \url{http://relativity.livingreviews.org/Articles/lrr-2010-4}

\bibitem[Will(2006)]{will2006confrontation}
Will, C.M.,
The Confrontation between General Relativity and Experiment, Living Reviews in Relativity 9, 3, 2006. Available at [2010/01/08]: \url{http://www.livingreviews.org/lrr-2006-3}

\bibitem[Willemenot(1997)]{willemenot1997pendule}
Willemenot, E., Pendule de torsion \`a suspension \'electrostatique, Ph.D. thesis, Universit\'e de Paris-Sud - Paris XI, 1997.

\bibitem[Willson \& Mordvinov(2003)]{willson2003secular}
Willson, R.C., \& Mordvinov, A.V.,
Secular total solar irradiance trend during solar cycles 21-23, Geophys. Res. Lett 30, 1199, 2003.

\bibitem[Wolf et al.(2009)]{wolf2009quantum}
Wolf P., Bord\'e C.J., Clairon A., et al.,
Quantum Physics Exploring Gravity in the Outer Solar System: The Sagas Project, Exp. Astron. 23, 651-689, 2009.

\bibitem[Zarrouati(1987)]{zarrouati1987trajectoires}
Zarrouati, O., Trajectoires spatiales, Cepadues Editions, Toulouse, France, 1987.

\end{thebibliography}
\end{document}